\documentstyle[12pt]{article}
\begin{document}
\newcommand{\beq}{\begin{equation}}
\newcommand{\eeq}{\end{equation}}
\newcommand{\ob}{\vert_{\partial M}=0}
\newcommand{\cB}{{\cal B}}
\begin{titlepage}
\title{Diffeomorphism invariant eigenvalue problem for
 metric perturbations in a bounded region}
\author{Valeri N. Marachevsky and Dmitri V. Vassilevich
\thanks{e.mail: vasilevich@phim.niif.spb.su}}
\date{ }
\maketitle
\centerline{\it Department of Theoretical Physics, St.Petersburg
University,}
\centerline{\it 198904 St.Petersburg, Russia \/}
\abstract{We suggest a method of construction of general
diffeomorphism invariant boundary conditions for metric
fluctuations. The case of $d+1$ dimensional Euclidean disk
is studied in detail. The eigenvalue problem for the Laplace
operator on metric perturbations is reduced to that on
$d$-dimensional vector, tensor and scalar fields. Explicit
form of the eigenfunctions of the Laplace operator is derived.
We also study restrictions on boundary conditions which are
imposed by hermiticity of the Laplace operator.}
\vfill
SPbU-IP-95
\end{titlepage}
\section{Introduction}

One of the main problems in quantum cosmology is to find
a suitable set of boundary conditions for graviton
perturbations (see monograph \cite{Esp} and review
\cite{BarR}). The contribution of the so called physical
degrees of freedom to the one-loop prefactor of the
wave function of the Universe was first evaluated by
Schleich \cite{Sch}. However, it is generally accepted now
that the contribution of other ("non-physical") components
of the metric perturbations does not cancel that of the
ghost fields \cite{D30,EKMP1,EKMP2,EK}. This means that
to define the one-loop prefactor one needs boundary
conditions for all components of the metric perturbations
and ghosts. One possible choice is the Luckock--Moss--Poletti
\cite{LMP} mixed boundary conditions. These boundary conditions
are local and they ensure hermiticity of the Laplace operator.
However, these boundary conditions are not completely gauge
invariant. Recently it was understood \cite{D41,EKMP2}
that for non totally geodesic boundary locality contradicts
gauge invariance. Hence the non-local gauge invariant
boundary conditions suggested by Barvinsky \cite{BarB}
look naturally. However, hermiticity of the Laplace operator
for these boundary conditions was not proved. The same is
true for another set \cite{EK} of non-local boundary
conditions. Local gauge invariant boundary conditions
were suggested \cite{D41} for $2d$ gravity with dynamical
torsion, but it is not clear whether this result can be
extended to higher dimensions.

    The purpose of the present work is to define most
general diffeomorphism invariant boundary conditions for
the metric perturbations, to find eigenfunctions of the Laplace
operator on a disk and to study restrictions imposed by
hermiticity of the Laplacian. We propose to impose boundary
conditions {\it independently\/} on gauge-fixed fields and
gauge transformations. This gives a large family of gauge
invariant boundary conditions (section 2).
Next we reduce systematically
the gauge-invariant eigenvalue problem for metric perturbations
to lower spin problems. For covariant gauge conditions on a
disk we find the eigenfunctions explicitly (sec. 3).
As a last step (sec. 4),
we consider the hermiticity condition for the Laplace operator
on transverse traceless metric perturbations. All boundary
conditions leading to hermitian Laplace operator are
specified. A part of this boundary conditions has very
unusual form. In Sec. 5 we discuss the results.

\section{General construction }

    The Einstein general relativity being formulated as a
theory of dynamical metric field exhibits the diffeomorphism
invariance. The infinitesimal diffeomorphism transformations
with the parameter $\xi_\mu$ act on metric fluctuations
$h_{\mu \nu}$ as follows:
\beq
h_{\mu \nu} \to h_{\mu \nu} +(L\xi )_{\mu \nu}, \quad
(L\xi )_{\mu \nu}=\nabla_\mu \xi_\nu +
\nabla_\nu \xi_\mu  \label{eq:dif} \eeq
where $\nabla$ is covariant derivative with respect to
background metric $g_{\mu \nu}$. Throughout this paper we
suppose that $g_{\mu \nu}$ is flat.

    Consider boundary conditions for $h$ :
\beq
\cB h \ob \label{eq:bc}
\eeq
with some boundary operators $\cB$. The
boundary conditions (\ref{eq:bc}) are diffeomorphism
invariant if there is a boundary operator $\cB_\xi$
such that
\beq
\cB L\xi \ob \label{eq:inv}
\eeq
provided $\xi$ satisfies
\beq
\cB_\xi \xi \ob \label{eq:bcxi}
\eeq
This means that diffeomorphism transformations map the
functional space defined by eq. (\ref{eq:bc}) onto itself.
For example, for the Barvinsky boundary conditions \cite{BarB}
the boundary operator is $\cB_\xi =1$, which gives Dirichlet
boundary conditions for $\xi$ and, consequently, for ghosts.

    One can construct general diffeomorphism invariant
boundary conditions in the following way. Consider a linear
background gauge
condition $\chi $. Denote the metric fluctuations satisfying
the equation $\chi (h)=0$ by $h^\perp$. For any $h$ there is
a unique decomposition
\beq
h=h^\perp +Lv \label{eq:perp} \eeq
with complementary
non-local projectors $P^\perp$ and $P^L$ such that
$h^\perp =P^\perp h$, $Lv=P^Lh$. The following boundary
operator
\beq
\cB h =\cB^\perp P^\perp h + \cB_\xi L^{-1}P^Lh
\label{eq:bcin}
\eeq
defines diffeomorphism invariant boundary conditions for
arbitrary $\cB^\perp$ and $\cB_\xi$. Note that $L$ is
invertible on $P^Lh$. In other words, the boundary conditions
(\ref{eq:bcin}) are imposed independently on gauge-fixed
components $h^\perp$ and pure gauge degrees of freedom.

    Up to this point, gauge condition $\chi$ was arbitrary.
We shall use the relativistic gauge condition
\beq
\chi (h)_\mu =\nabla^\nu h_{\mu \nu}-\alpha \nabla_\mu
h^\nu_\nu \label{eq:relg}
\eeq
with a real parameter $\alpha$. This choice is most convenient
because the Laplace operator $\Delta =\nabla^\mu \nabla_\mu$
commutes with the projectors $P^\perp$ and $P^L$, and
\beq
\Delta : \{ h^\perp\} \to \{ h^\perp\} ,\quad
\Delta Lv =L\Delta v \label{eq:com}
\eeq
This means that one can consider separately the eigenvalue
problem for the Laplace operator on the spaces $h^\perp$
and $Lv$. The Laplace operator in the latter space is reduced
to the vector Laplace operator, which was studied in literature
in detail
(see e.g \cite{Esp,LMP}). Furthermore, $h^\perp$ can be decomposed in
transverse traceless fluctuations $h^{TT}$ and a part
depending on a scalar field:
\beq
h^\perp_{\mu \nu} =h^{TT}_{\mu \nu} + h^\perp (\sigma ), \quad
h^\perp (\sigma )=
\left ( g_{\mu \nu} \frac {(1-\alpha )\Delta }{\alpha (d+1) -1} +
\nabla_\mu \nabla_\nu \right )\sigma
\label{eq:rasl}
\eeq
$h^{TT\mu}_\mu =0$, $\nabla^\mu h^{TT}_{\mu \nu}=0$. The
decomposition (\ref{eq:rasl}) commutes with the Laplace
operator:
\beq
\Delta : \{ h^{TT}\} \to \{ h^{TT}\} ,\quad
\Delta h^\perp (\sigma ) =h^\perp (\Delta \sigma ) \label{eq:com1}
\eeq
Thus the eigenvalue problem for the Laplace operator on $h^\perp$
is reduced to the eigenvalue problem on $h^{TT}$ and relatively
simple scalar eigenvalue problem.

    In the following section we shall study the eigenvalue
equation for the Laplace operator on $h^{TT}$.

\section{Transverse traceless metric perturbations on a disk}

Consider $d+1$ dimensional unit disk $M$ with the metric
\beq
ds^2=dr^2 +r^2 d\Omega^2, \quad 0\le r \le 1 ,
\label{eq:3met}
\eeq
where $d\Omega^2$ is the metric on unit sphere $S^d$.
We shall need explicit expressions for the Laplace operator
acting on  scalar field $\phi$, vector  $A_\mu$ and
rank two symmetric tensor $h_{\mu \nu}$:
\begin{eqnarray}
\Delta \phi & = &
(\partial^2_0+\frac dr \partial_0+^{(d)}\Delta )\phi
\label{eq:3sc} \\
(\Delta A)_0 & = & (\partial_0^2 +\frac dr \partial_0
+^{(d)}\Delta - \frac d{r^2} )A_0 -\frac 2r ^{(d)}\nabla^i
A_i, \nonumber \\
(\Delta A)_i & = & (\partial_0^2 + \frac {d-2}r \partial_0
+^{(d)}\Delta -\frac {d-1}{r^2} )A_i +\frac 2r \partial_iA_0
\label{eq:3ve} \\
(\Delta h)_{00} & = &
(\partial^2_0+\frac dr \partial_0+^{(d)}\Delta -\frac {2d}{r^2})
h_{00} - \frac 4r\ ^{(d)}\nabla_i h_0^i
+\frac 2{r^2}h_i^i \nonumber \\
(\Delta h)_{0i} & = &
(\partial^2_0+\frac {d-2}r \partial_0+^{(d)}\Delta
-\frac {2d+1}{r^2})h_{0i} +\frac 2r \partial_i h_{00}
-\frac 2r\ ^{(d)}\nabla^kh_{ik} \nonumber \\
(\Delta h)_{ik} & = &
(\partial^2_0+\frac {d-4}r \partial_0+^{(d)}\Delta
-\frac {2d-4}{r^2})h_{ik} + \nonumber \\
\ & \ & +\frac 2r (\ ^{(d)}\nabla_i h_{0k}+^{(d)}\nabla_k
h_{0i})+ \frac 2{r^2}g_{ik} h_{00}
\label{eq:3te}
\end{eqnarray}
where $\ ^{(d)}\nabla$ and $\ ^{(d)}\Delta$ are the
$d$-dimensional covariant derivative and the Laplace
operator with respect to $d$-dimensional metric $g_{ik}$,
respectively.

Let us expand $h_{00}$, $h_{0i}$ and $h_{ik}$ in sums
of irreducible harmonics on $S^d$:
\begin{eqnarray}
h_{00} & = & s_1 \nonumber \\
h_{0i} & = & \partial_i s_2 + u_{1,i} \label{eq:3ex} \\
h_{ik} & = & \ ^{(d)}\nabla_i u_{2,k} +^{(d)}\nabla_k u_{2,i}
+(\ ^{(d)}\nabla_i \ ^{(d)}\nabla_k-\frac 1d g_{ik}
\ ^{(d)}\Delta )s_3 \nonumber \\
 &\ & -\frac 1d g_{ik} s_1 +h_{ik}^{tt} \nonumber
\end{eqnarray}
Here $s_A$ are scalar fields, $u_{A,i}$ are transversal
vectors, $\ ^{(d)}\nabla^iu_{A,i}=0$, and $h^{tt}$ is
transversal traceless tensor, $h^{tt}_{ij}g^{ij}=0$,
$\ ^{(d)}\nabla^ih^{tt}_{ik}=0$. In the expansions
(\ref{eq:3ex}) we used the fact that the $d+1$ dimensional
trace of $h_{\mu \nu}$ is zero. Up to some redefinitions
the representation (\ref{eq:3ex}) is the same as was used
by other authors \cite{EKMP1}. Here however we shall
apply it to transversal fields only.
The $d+1$ dimensional
transversality condition, $\nabla^\mu h_{\mu \nu}=0$,
reads:
\begin{eqnarray}
\nu =0: &\quad & 0=^{(d)}\nabla^i h_{0i} +(\partial_0 +\frac dr )
h_{00} -\frac 1r h^i_i \nonumber \\
\nu =i: &\quad & 0=^{(d)}\nabla^j h_{ji} +
(\partial_0 +\frac dr )h_{0i} \label{eq:3tra}
\end{eqnarray}
Due to the equations (\ref{eq:3tra}) the fields $s_A$ and $u_A$
can be expressed via single independent function of each kind.
General solution of (\ref{eq:3tra}) has the following form:
\begin{eqnarray}
h^{TT}_{\mu \nu}& = & h^{tt}_{\mu \nu} +h_{\mu \nu}(u_i)
+h_{\mu \nu}(s) \label{eq:3gen} \\
h_{0i}(u) & = & -(\ ^{(d)}\Delta +\frac {d-1}{r^2} )ru_i
\nonumber \\
h_{ik}(u) & = & r(\partial_0 +\frac {d-1}r )
(\ ^{(d)}\nabla_i u_k+^{(d)}\nabla_k u_i) \label{eq:3solu}\\
h_{00}(s) & = & (d-1)^{(d)}\Delta (\ ^{(d)}\Delta +
\frac d{r^2}) r^2 s ; \nonumber \\
h_{0i}(s) & = & -(d-1)\partial_i (\partial_0 +\frac {d-1}r )
(\ ^{(d)}\Delta +
\frac d{r^2}) r^2 s ; \nonumber \\
h_{ik}(s) & = & (\ ^{(d)}\nabla_i\ ^{(d)}\nabla_k -\frac 1d
g_{ik}\ ^{(d)}\Delta ) (d\partial_0^2 +\frac {2d^2-5d}r
\partial_0 +\frac {d(d-3)^2}{r^2} \nonumber \\
\ & \ &
+^{(d)}\Delta )
r^2s
 -\frac {d-1}d g_{ik}\ ^{(d)}\Delta (\ ^{(d)}\Delta +
\frac d{r^2}) r^2 s .\label{eq:3sols}
\end{eqnarray}
The components $h^{tt}_{00}$, $h^{tt}_{0i}$, $h_{00}(u)$
vanish identically.

Using the explicit expressions (\ref{eq:3sc}), (\ref{eq:3ve})
and (\ref{eq:3te}) one can demonstrate by straightforward
computations that
\beq
\Delta h_{\mu \nu}(u_i) =h_{\mu \nu}(\Delta u_i),
\quad
\Delta h_{\mu \nu}(s) =h_{\mu \nu}(\Delta s).
\label{eq:3lap}
\eeq
We see, that the eigenfunctions of the Laplace
operator $\Delta$ on the space of transverse
traceless tensors can be defined through
the eigenfunctions of the same operator on
the so called physical components $s$, $u$ and
$h^{tt}$ of scalar, vector and tensor fields.
The spectrum of the Laplace operator on the latter
components is well known (see e.g. monograph
\cite{Esp}).
Let us represent $h^{tt}$, $u$ and $s$ as Fourier
series of hyperspherical harmonics on $S^d$:
\begin{eqnarray}
h^{tt}_{ik}(x^\mu ) & = &
\sum_{(l)} H_{(l)}(r) Y^{tt(l)}_{ik} (x^i) , \nonumber \\
 u_k(x^\mu ) & = &
\sum_{(l)} U_{(l)}(r) Y^{u(l)}_{k} (x^i) , \nonumber \\
s(x^\mu ) & = &
\sum_{(l)} S_{(l)}(r) Y^{s(l)} (x^i) . \label{eq:3Fo}
\end{eqnarray}
The spectrum of the $d$-dimensional Laplace operator on
these harmonics is well known \cite{RO}:
\begin{eqnarray}
\ ^{(d)}\Delta Y^{s(l)}(x_i) & = &
-\frac 1{r^2} l(l+d-1)Y^{s(l)}(x_i) , \quad l \ge 0
\nonumber \\
\ ^{(d)}\Delta Y^{u(l)}_k(x_i) & = &
-\frac 1{r^2} [ l(l+d-1)-1] Y_k^{u(l)}(x_i) ,\quad l \ge 1
\nonumber \\
\ ^{(d)}\Delta Y^{tt(l)}_{jk}(x_i) & = &
-\frac 1{r^2} [l(l+d-1)-2]Y^{tt(l)}_{jk}(x_i)
\quad l \ge 2
\label{eq:3spec}
\end{eqnarray}
The corresponding degeneracies are
\begin{eqnarray}
D_l^s & = &\frac {(2l+d-1)(l+d-2)!}{l!(d-1)!} \nonumber \\
D_l^u & = &\frac {l(l+d-1)(2l+d-1)(l+n-3)!}{(d-2)!(l+1)!}
\nonumber \\
D_l^{tt} & = &\frac {(d+1)(d-2)(l+d)(2l+d-1)(l+d-3)!}{2(d-1)!
(l+1)!} \label{eq:3deg}
\end{eqnarray}
Note, that the $d+1$-dimensional Laplace operators
(\ref{eq:3ve}) and (\ref{eq:3te}) map the spaces of
$d$-dimensional transverse vectors and transverse
traceless tensors onto themselves. The eigenvalue
equations for the $d+1$ dimensional Laplace operator
$\Delta$ are reduced to ordinary differential equations
for the functions $H_{(l)}$, $U_{(l)}$ and $S_{(l)}$.
One can easily find the eigenfunctions:
\begin{eqnarray}
s_{l,\lambda} & \propto & r^{(1-d)/2}J_{(d-1)/2+l}(r\lambda )
Y^{s(l)}(x^i) , \nonumber \\
u_{l,\lambda k} & \propto & r^{(3-d)/2}J_{(d-1)/2+l}(r\lambda )
Y^{u(l)}(x^i) , \nonumber \\
h^{tt}_{l,\lambda k} & \propto & r^{(5-d)/2}J_{(d-1)/2+l}(r\lambda )
Y^{tt(l)}(x^i), \label{eq:3eig} \\
& \ & l=2,3,4,... \nonumber
\end{eqnarray}
The values $l=0,1$ are excluded because the corresponding
 harmonics
generate zero modes of the mappings $s\to h^{TT}$ and
$u\to h^{TT}$. The eigenvalues $-\lambda^2$ of the Laplace
operator $\Delta$ are defined by boundary conditions.
Degeneracies of each eigenvalue are given by (\ref{eq:3deg}).

    In order to find gauge invariant boundary conditions
for the metric fluctuations on a disk one should choose
some (arbitrary) boundary conditions for the fields $v$,
$\sigma$, $s$, $u$ and $h^{tt}$ entering the decompositions
(\ref{eq:perp}), (\ref{eq:rasl}) and (\ref{eq:3gen}). The
equations (\ref{eq:com}), (\ref{eq:com1}), (\ref{eq:3lap}),
(\ref{eq:3eig}) define the spectrum of the Laplace operator.

\section{Selfadjointness analysis}

Let us find the boundary conditions for $h^{TT}$
which lead to selfadjoint Laplacian. We shall use
ordinary inner product without surface terms in the
space of rank two symmetric tensor fields:
\beq
<h',h>=\int d^{d+1}x \sqrt g g^{\mu \nu}
g^{\eta \kappa} {h'}_{\mu \eta} h_{\nu \kappa}
\label{eq:4inn}
\eeq
The selfadjointness means that{\footnote{Strictly speaking,
this equation means that the Laplacian is symmetric.
For selfadjointness, one should also require that an operator
and its ajoint have coinciding domains of definition.}}
 $<h',\Delta h>=<\Delta h',h>$ provided $h'$ and $h$ satisfy
some boundary conditions. This is equivalent to vanishing
of the following surface integral
\beq
\int_{\partial M} g^{\mu \nu}
g^{\eta \kappa} ({h'}_{\mu \eta} \partial_0 h_{\nu \kappa}-
h_{\mu \eta} \partial_0 {h'}_{\nu \kappa})=0
\label{eq:4sur}
\eeq
Due to orthogonality of the tensor harmonics on $S^d$ the
equation (\ref{eq:4sur}) should be satisfied by all fields
$s$, $u$ and $h^{tt}$ and all values of $l$ in the
decompositions (\ref{eq:3sols}) and (\ref{eq:3Fo})
independently. Let us suppose that the boundary conditions
are $SO(d)$ invariant.
Thus the equation (\ref{eq:4sur})
generates some restrictions on boundary conditions
for the coefficient functions $H_{(l)}$, $U_{(l)}$ and
$S_{(l)}$ for any $l$.

    Consider first the $h^{tt}$ fluctuations. The integral
(\ref{eq:4sur}) vanishes if $H_{(l)}$ satisfies one of the
following boundary conditions:
\beq
H_{(l)}\ob \quad {\rm or} \quad
(\partial_0 +C_l^{tt})H{(l)}\ob
\label{eq:4tt}
\eeq
with arbitrary constants $C_l^{tt}$. The boundary conditions
(\ref{eq:4tt}) are very general. This means that the boundary
conditions for the so called physical degrees of freedom
are not restricted by the invariance or selfadjointness
requirements and are defined by physics only.

    Consider now the $h(u)$ fluctuations. One can check the
equation (\ref{eq:4sur}) on the eigenfunctions (\ref{eq:3Fo}),
(\ref{eq:3eig}). Namely, we can suppose that $u$ and $u'$
correspond to some fixed eigenvalues of $\Delta$ and
$r^2\ ^{(d)}\Delta$. Denote the corresponding quantum numbers
as $\lambda$, $l$, $\lambda '$ and $l'$. Due to orthogonality
of spherical harmonics the integral in (\ref{eq:4sur})
vanishes identically for $l'\ne l$. Hence we put $l'=l$.
Consider first the Neumann boundary conditions for $U_{(l)}$:
\beq
(\partial_0 +C^u_l )U_{(l)} \ob \label{eq:4bcu}
\eeq
To preserve the $SO(d)$ invariance of the boundary conditions
$C^u_l$ should be constant. By substituting (\ref{eq:4bcu})
in (\ref{eq:4sur}) we obtain the following condition
\beq
(\lambda^2 -{\lambda '}^2)(C_l^u-d+1)(l(l+d-1)-d)
\int_{\partial M} d^d x \sqrt {\ ^{(d)}g} u^j {u'}_j=0
\label{eq:4cond}
\eeq
Since $\lambda$ and $\lambda '$ have, in general, different
values, the condition (\ref{eq:4cond}) implies that
$C_l^u=d-1$. One can also check that (\ref{eq:4sur}) holds
if $u_i$ satisfies Dirichlet boundary conditions. Thus to
give selfadjoint Laplace operator, $U_{(l)}$ should satisfy
one of the two following boundary conditions:
\beq
U_{(l)}\ob \quad {\rm or} \quad
(\partial_0 +d-1)U_{(l)} \ob \label{eq:4u}
\eeq

    The case of $h(s)$ can be analyzed along the same lines.
We obtain the Laplace operator is selfadjoint only for
Neumann boundary conditions:
\begin{eqnarray}
(\partial_0 +C_l^s )S_{(l)} \ob \nonumber \\
C_l^s=d-1+
\frac 1d (\Lambda_l \pm \sqrt{\Lambda_l^2-d\Lambda_l})
\label{eq:4s} \\
\Lambda_l =l(l+d-1) \nonumber
\end{eqnarray}
where $\Lambda_l$ are eigenvalues of $-r^2\ ^{(d)}\Delta$.
These boundary conditions (\ref{eq:4s}) are non-local and
even non polynomial in tangential derivatives. However,
due to the fact that the conditions (\ref{eq:4s}) contain
only first order normal derivatives, they define a well
posed boundary problem for the Laplace operator. One can see
this directly by using the decomposition (\ref{eq:3Fo}) and
analyzing ordinary differential equations for $S_{(l)}$.
This completes our analysis of boundary conditions for $h^{TT}$
leading to symmetric Laplace operator.

    Strictly speaking, to use spectral representation for the
Laplace operator in the path integral one should demonstrate
selfadjointness. This might be done by using one of the
criteria of monograph \cite{RS}. We postpone this tedious
task to a future publication where we shall analyze path
integral over metric perturbations. In any case, for boundary
conditions other than listed here the Laplace operator is not
symmetric and, hence, is not selfadjoint.

    Boundary conditions for other components of the metric
fluctuations can be analyzed along the same lines. Consider the
gauge (\ref{eq:relg}) with $\alpha \to (d+1)^{-1}$. Instead of
$h^\perp (\sigma )$ it is convenient to introduce $h^\perp
(\omega )=g_{\mu \nu} \omega (x)$. One can easily demonstrate
that the surface integral (\ref{eq:4sur}) vanishes for
Dirichlet boundary conditions as well as for Neumann boundary
conditions with arbitrary constant. Hence at least in this
gauge selfajointness does not mean any restrictions on the
trace part of metric fluctuations. In this section we will not
consider the pure gauge part $Lv$, because this fluctuations
represent zero modes of the gravitational action.

\section{Discussion and conclusions}

    In order to describe a physical theory the boundary
conditions for metric fluctuations should satisfy some
requirements. Let us discuss them in brief.

    (i) Some part of boundary conditions is fixed by
{\it physics}. Usually, these are the boundary conditions
for the so called physical degrees of freedom, which
correspond to $h^{tt}$ in our notations. Fortunately,
these boundary conditions are not fixed by selfconsistency
conditions considered in the present paper.

    (ii) {\it Diffeomorphism invariance}. This is the basic
invariance of gravitation. This property should be preserved
to obtain meaningful quantum theory. If boundary conditions
are not gauge invariant, than, for example, on-shell effective
action becomes gauge-dependent \cite{DV5}. One can satisfy
the diffeomorphism invariance in a relatively simple way.
One can choose the condition (\ref{eq:bcin}) with arbitrary
conditions for the gauge-fixed part $h^\perp$ and gauge
transformations. Furthermore, the eigenvalue problem for
the Laplace operator on metric perturbations is reduced
to eigenvalue problems for $h^{tt}$ and several $d$-dimensional
vectors and scalars. On a disk, explicit form of the
eigenfunctions was obtained.

    (iii) The Laplace operator should be {\it Hermitian \/} with
respect to inner product (\ref{eq:4inn}) without surface terms.
Only if an operator $K$ is Hermitian with respect to an inner
product $<\ ,\ >$, the path integral
\beq
\int {\cal D} \phi \exp (-<\phi ,K\phi >)
\label{eq:5int}
\eeq
is defined by product of eigenvalues of $K$. We considered
the Laplace operator and the inner product (\ref{eq:4inn})
because, first, they are the most simple choice, and second,
because this construction is used in all actual computations,
though sometimes implicitly. The boundary conditions
(\ref{eq:4s}) for $h^{TT}(s)$ are somewhat unusual.
May be these boundary conditions do nevertheless admit
a nice geometric interpretation. It could happen as well
that a more careful consideration of boundary terms
in the action will modify drastically the results of
the previous section.

    Unfortunately, complicated form of the boundary
conditions (\ref{eq:4s}) prevents us from using powerful
methods of evaluation of functional determinants
designed recently \cite{D36,BGV,BKKM,Eli,Dow}.

    Another way to determine proper boundary conditions
may be to introduce arbitrary surface coupling and look
for the renormalization group fixed points in the sense
of recent works \cite{Odin}.

\section*{Acknowledgments}

The authors would like to thank Ivan Avramidi, Giampiero
Esposito and Alexander Kamenshchik for discussions. This
work was supported by the Russian Foundation for Fundamental
Studies, and by St.Petersburg Regional
Program through GRACENAS, project M94-2.2$\Pi$-18.

    \end{document}